\documentclass{article}
\usepackage{amsmath,amsthm}
\usepackage{amssymb,latexsym}
\usepackage[mathscr]{eucal}
\usepackage{setspace}
\usepackage{graphics}
\usepackage{array}

\DeclareMathSizes{13}{13}{9}{8}

\newcommand{\rb}{\right]}
\newcommand{\lb}{\left[}

\newcommand{\beq}{\begin{equation}}
\newcommand{\eq}{\end{equation}}

\usepackage{graphicx}
\graphicspath{{converted_graphics/}}
\begin{document}

\title{Unsteady Hall Magnetohydrodynamics near a Hyperbolic Magnetic Neutral Line: An Exact Solution}         
\author{Bhimsen K. Shivamoggi\footnote{\large Permanent Address: University of Central Florida, Orlando, FL 32816-1364}\\
Los Alamos National Laboratory\\
Los Alamos, NM 87545}        
\date{}          
\maketitle

\large{\bf Abstract-} Unsteady Hall Magnetohydrodynamics (MHD) near a hyperbolic magnetic neutral line  is investigated. An exact analytical solution describing a self-similar evolution is given. This solution shows a negligible impact on the current-sheet formation process near the hyperbolic magnetic neutral line at small times by the Hall effect but, subsequently, a quenching by the Hall effect of the finite-time singularity exhibited in ideal MHD and, hence a prevention of the current density blow-up at large times. The asymptotic result given by this time-dependent solution is in full quantitative agreement with the formulation of \textit{steady} Hall MHD near a $X$-type magnetic neutral line (Shivamoggi [23]). The latter formulation showed that this asymptotic result indeed corresponds to a hyperbolic configuration of the magnetic field lines in the \textit{steady} case.
\vspace{0.05in}

\section{Introduction}

\indent When a plasma collapses near the neutral line of the applied magnetic field the continual accumulation of the magnetic flux in the region of the neutral sheet puts the current sheet in a non-stationary state (Syrovatskii [1]). An exact self-similar solution of the MHD equations for a time-dependent, two-dimensional (2D) flow of an incompressible plasma in a hyperbolic magnetic field was given by Uberoi [2] and Chapman and Kendall [3]. This solution had an initialy current-free magnetic field, so it is not appropriate for the reconnection problem. This solution was modified (Shivamoggi [4]) so as to relax this restriction, and hence make it suitable for the reconnection problem. This solution was generalized to incorporate a uniform shear-strain rate in the plasma flow (Shivamoggi [5], Rollins and Shivamoggi [6]), so the magnetic field lines now undergo not only sweeping but also shearing by the plasma flow. The above solution predicted a sequence of events associated with the evolution of a current sheet in a hyperbolic magnetic field, in agreement with laboratory experiments (Frank [7]) on the collapse of a plasma near the hyperbolic neutral line. In recognition of the numerical results (Brunnel et al. [11]) showing the significant effect plasma density variations near the magnetic neutral point have on the magnetic reconnection processes taking place there the above solution was generalized further to incorporate density variations in the plasma (Rollins and Shivamoggi [12]). The current-sheet formation process was found to speed up in the presence of a plasma density build-up near the current sheet.

\indent Fast magnetic reconnection processes in laboratory (ex: sawtooth collapse in tokamak dischanges) and space (ex: solar flares and magnetospheric substorms) can be described using collisionless plasma models (Yamada et al. [13], and Shibata [14]). In a high-$\beta$ collisionless plasma, on length scales shorter than the ion skin depth $d_i$, the electrons decouple from the ions and the electron dynamics is governed by Hall currents (Sonnerup [15]).  The decoupling of ions and electrons in a narrow region around the magnetic neutral point allows for rapid electron flows in the ion dissipation region and hence a faster magnetic reconnection process (Mandt et al. [16] and Biskamp et al. [17]). 

\indent It may be mentioned that even in the absence of the Hall currents fast collisionless magnetic reconnection has been shown to be feasible (Scholer et al. [18], Jaroschek et al. [19], Bessho and Bhattacharjee [20], Ishizawa and Horiuchi [21]). This is caused by - 
\begin{itemize}
  \item inductive electric fields generated by lower hybrid drift instability, or
  \item non-gyrotropic pressure tensor effects caused by the ion-meandering motion near the magnetic neutral point.
\end{itemize}

\indent In recognition of the important role played by the Hall effect in fast magnetic reconnection processes an investigation of unsteady Hall MHD near a hyperbolic magnetic neutral line\footnote{\large This problem has also recently been considered by Litvinenko [22]; however, his solution is rather different from the present solution and does not show the asymptotic connection with \textit{steady} Hall MHD near the $X$-type magnetic neutral line.}  is therefore in order - this is the objective of this paper. The asymptotic result given by the time-dependent solution in question turns out to be in full quantitative agreement with the formulation of \textit{steady} Hall MHD near a $X$-type magnetic neutral line (Shivamoggi [23]).
\vspace{0.05in}

\section{Governing Equations for Hall MHD} 

\indent Consider an incompressible, two-fluid, quasi-neutral plasma. The  equations governing this plasma dynamics are (in usual notation) -

\beq 
nm_{e}[\frac{\partial \textbf{v}_{e}}{\partial t} + (\textbf{v}_{e} \cdot \nabla)\textbf{v}_{e}] = -\nabla p_{e} -ne (\textbf{E} + \frac {1} {c} \textbf{v}_{e}\times \textbf{B}) + ne \eta \textbf{J}
\eq
\begin{equation}
nm_{i} [\frac {\partial \textbf{v}_i}{\partial t} + (\textbf{v}_{i} \cdot \nabla) \textbf{v}_{i}] = -\nabla p_{i} + ne (\textbf{E} + \frac {1}{c} \textbf{v}_{i} \times \textbf{B}) - ne\eta \textbf{J}
\end{equation}

\begin{equation}
\nabla \cdot \textbf{v}_{e} = 0
\end{equation}

\begin{equation}
\nabla \cdot \textbf{v}_{i} = 0
\end{equation}

\begin{equation}
\nabla \cdot \textbf{B}   = 0
\end{equation}

\begin{equation}
\nabla\times \textbf{B}=\frac{1}{c} \textbf{J}
\end{equation}

\begin{equation}
\nabla \times \textbf{E}   =  -\frac{1}{c}\frac{\partial \textbf{B}}{\partial t}
\end{equation}

\noindent where,

\beq \textbf{J} \equiv ne (\textbf{v}_i - \textbf{v}_e).
\eq

Neglecting electron inertia ($m_e \Rightarrow 0$),~equations (1) and (2) can be combined to give an ion-fluid equation of motion -

\beq nm_i \large[\frac {\partial \textbf{v}_i}{\partial t} + (\textbf{v}_i \cdot \nabla)\textbf{v}_i] = - \nabla (p_e + p_i) + \frac {1} {c} \textbf{J}\times \textbf{B}
\eq

\noindent and a generalized Ohm's law - 

\beq \textbf{E} + \frac {1} {c} \textbf{v}_i \times \textbf{B} = \eta \textbf{J} + \frac {1} {nec}\textbf{J} \times \textbf{B}.
\eq

Non-dimensionalize distance with respect to a typical length scale $a$, magnetic field with respect to a typical magnetic field strength $B_0$, time with respect to the reference Alfv\`{e}n time $\tau_A \equiv \frac {a}{V_{A_0}}$ where $V_{A_0} \equiv \frac {B_0}{\sqrt\rho}$ , $\rho\equiv m_i n,$ and introduce the magnetic stream function according to 

\beq \textbf{B} = \nabla \psi \times \mathbf{\hat i}_z + b~ \mathbf{\hat i}_z
\eq
and write the ion-fluid velocity as

\beq \textbf{v}_i = (\mathbf{\hat i}_z \times \textbf{v}_i)\times \mathbf{\hat i}_z + w \mathbf{\hat i}_z \equiv \textbf{v} + w \mathbf{\hat i}_z \eq

\noindent and assume the physical quantities of interest have no variation along the $z-$direction. Equations (9) and (10) then yield

\beq [\frac{\partial}{\partial t} + (\textbf{v} \cdot \nabla)]\textbf{v} = -\nabla P - (\nabla^2 \psi) \nabla\psi
\eq

\beq \large[\frac{\partial}{\partial t} + (\textbf{v} \cdot \nabla)] \psi + \sigma [b, \psi] = \hat{\eta} \nabla^2 \psi 
\eq

\beq [\frac{\partial}{\partial t} + (\textbf{v} \cdot \nabla)] b + \sigma \large [\psi, \nabla^2\psi] + [\psi, w] = \hat{\eta} \nabla^2 b
\eq

\beq [\frac{\partial}{\partial t} + (\textbf{v} \cdot \nabla)] w -  [b, \psi] = 0
\eq

\noindent where, 

$$
[A,B] \equiv \nabla A \times \nabla B \cdot {\hat{\textbf{i}}_z},~ P \equiv p_e + p_i + b^2,~  \sigma \equiv \frac {d_i}{a}, ~ \hat{\eta} \equiv \frac {\eta c^2 \tau_A}{{a}^2}.
$$
\vspace{0.05in}

\section{Hall MHD Near a Hyperbolic Magnetic Neutral Line}

\indent Consider the initial-value problem near a hyperbolic magnetic neutral line in Hall MHD with initial conditions - 

\begin{equation} 
t = 0 : v_x = -\gamma_0 x,~~  v_y = \gamma_0 y, ~~  w = - 1/\sigma (kx^2 - y^2),~~  \psi = k x^2 - y^2, ~~  b = C x y
\end{equation}

\noindent where $\gamma_0$ and $k$ are externally determined parameters with $\gamma_0 > 0$ and $C > 0$. This initial condition describes a stagnation-point plasma flow impinging transversely onto the $x=0$ plane and incorporates equations (4) and (5). The spatial structure for the out-of-plane magnetic field described by this initial condition is in recognition of the \textit{quadrupolar} out-of-plane magnetic field $b$ pattern characterizing the Hall effects (Terasawa [24]). Laboratory experiments (Ren et al. [25]) and \textit{in situ} measurements in the magnetotail (Fujimoto et al. [26], Nagai et al. [27], Oieroset et al. [28]) have also confirmed the latter signature of the Hall effect. The Hall magnetic field $b$ is believed to be produced by the dragging of the in-plane magnetic field in the out-of-plane direction by the electrons near the $X$-type magnetic neutral line ([16]).

The Lorentz force due to the initial magnetic field is 

\beq t = 0 : \textbf{J} \times \textbf{B} = 4(1-k) k x \mathbf {\hat{i}}_x - 4 (1 - k) y \mathbf{\hat i}_y - 2C(k x^2 + y^2) \mathbf{\hat i}_z.
\eq

\noindent We take $k > 1$       so that this Lorentz force is directed so as to maintain the prescribed initial stagnation-point flow.

Let us assume that the solution, for $t > 0$,   of equations (4), (5) and (13) - (16) with the above initial conditions is of the self-similar form 

\beq v_x (x,y,t) = - \dot{\gamma} (t) x,~  v_y(x, y, t) = \dot{\gamma} (t) y\notag
\eq
\beq w(x, y, t)= \frac{1}\sigma [\beta (t) y^2 - k \alpha(t)x^2]\notag
\eq
\beq \psi (x, y, t) = k \alpha (t) x^2 - \beta (t) y^2\notag
\eq
\beq b (x, y, t) = C x y\notag
\eq
\beq P (x, y, t) = -\frac{1}{2}\nu (t)(x^2 +y^2)+ P_0,~\nu (t)> 0
\eq

\noindent with 

\beq t = 0 : \alpha = \beta = 1,\dot{\gamma} = \gamma_0.
\eq

For the solution (19), $\nabla^2\psi = f (t)$ and $ \nabla^2 b = 0$, so the effect of resistivity in this case is to add a function of $t$ to $\psi$ (which leaves the magnetic field unaltered) and hence to introduce an electric field along the z-axis. We therefore drop the resistivity in the following. Further, for an incompressible plasma, the pressure does not have a dynamical role. So, it is forced to be an enslaved variable in the sense that its form is chosen so as to be compatible with equations (13) - (16) given the ansaetze for $v_x, v_y, w, \psi $ and $b$.\footnote{\large In the generic situation, it may be mentioned that there are six parameters - two for the magnetic flux function $\psi$, one for the out-of-plane magnetic field $b$, and three for the velocity field $(\textbf{v}, w)$, to be determined by only five scalar equations (13)-(16), so there is non-uniqueness in the solution. This is resolved by specifying the parameters as in (19) to close the system, so (19) is \textit{one} exact solution. However, this exact solution turns out to have considerable physical significance, as discussed in the following.} \footnote{\large It may be noted that Litvinenko [22] sets up the solution and the inital conditions for $w$ and $b$ rather different from those prescribed in (19) and (20). In Litvinenko's solution the amplitude of the driving plasma flow velocity field is taken to be time indepentent, i.e., $\dot\gamma = const$, while in the present solution the out-of-plane magnetic field is taken to be time independent, i.e., $b =  Cxy, C = const$. This difference turns out to lend considerable physical significance to the present solution.}

Substituting (19) into equation(13), we obtain \beq \ddot{\gamma} = 2 (k^2\alpha^2 - \beta^2)\eq

\noindent while equation (14) gives

\beq\dot{\alpha} - 2 (\dot{\gamma} + \sigma C)\alpha = 0\eq

\beq\dot{\beta} + 2 (\dot{\gamma} + \sigma C)\beta = 0\eq

\noindent Equations (15) and (16) are identically satisfied by the solution (19).

We have from equations (21)-(23),
\beq\alpha(t) = e^{2(\gamma + \sigma C t)}
\eq

\beq\beta (t) = e^{-2(\gamma +\sigma C t)}
\eq

\beq \dot{\gamma}^2 + 2 \sigma C \dot{\gamma} =  ~\large [k^2 e^{4(\gamma + \sigma C t)} + e^{-4(\gamma + \sigma C t)}] - A
\eq

\noindent with

\beq t = 0 : \gamma = 0.\eq

Equation (26) along with (20) and (27) yields
\beq
A = (k^2 + 1) - \gamma_0^2 -2 \sigma C \gamma_0.
\eq

\noindent It may be noted that (24) and (25) are consistent with the ion-fluid incompressibility 
condition - 

\begin{equation}
\frac {d}{dt}\left(\frac{1}{2} \ell n \lb \alpha(t)\beta(t)\rb\right) =\nabla\cdot \textbf {v}= 0
\end{equation}

\vspace{0.15in}

\noindent which is derivable from equation (14) on substituting (19).

\indent For small $t$, equation (26) gives \beq \gamma(t) \approx \gamma_0 t + (k^2 -1) t^2\eq

\noindent while for large $t$, equation (26) gives \beq\tag{31 a,b}\gamma (t)~ \approx  -\sigma Ct~ \text{or}~ \dot{\gamma}(t)\approx -\sigma C,~ t ~large\eq

(30) shows that the Hall effect $(\sigma \neq 0)$ does not materialize to $O(t^2)$. However,~(31) shows that the finite-time singularity exhibited in ideal MHD (Shivamoggi [29]) is quenched by the Hall effect. Thus, though the Hall effect does not impact the current-sheet formation process for small times, it prevents the current density blow-up at large times. This result may be further appreciated by noting that equations (24)-(26) lead to the \textit{exact} invariant - 
\beq \tag{32}(\dot{\gamma} + \sigma C)^2 - (k^2 \alpha^2 + \beta^2) = const. \eq
\noindent (32) clearly shows the suppression of the plasma collapse process near a hyperbolic magnetic neutral line in Hall MHD, for large times (when (31b) becomes valid). Physically, the suppression of the plasma collapse process near a hyperbolic magnetic neutral point in Hall MHD appears to be caused by the dispersive activity of whistler waves which is known to lead to current-sheet broadening, as confirmed by laboratory experiments (Urrutia et al. [30]) and satellite observations at the magnetopause and the magnetotail plasma sheet (Sonnerup et al. [31], Fairfield et al. [32]) as well as numerical simulations (Shay et al. [33] and [34]). (31b), in conjunction with (19), also shows that, for large $t$, the level curves of the out-of-plane magnetic field are also the streamlines of the in-plane ion flow. 
\noindent  It is pertinent to note that the asymptotic result (31b) is in full quantitative agreement with the  formulation of \textit{steady} Hall MHD near a $X$-type magnetic neutral line ([23]). The latter formulation showed that (31b) indeed corresponds to a hyperbolic configuration of the magnetic field lines in the \textit{steady} state.

\section{Discussion}

\indent In recognition of the important role played by the Hall effect in fast magnetic reconnection processes this paper makes an investigation of unsteady Hall MHD near a hyperbolic magnetic neutral line in Hall MHD. The Hall effect is found not to impact the current-sheet evolution  process near the hyperbolic magnetic neutral line for small times. However, subsequently, the Hall effect is found to quench the finite-time singularity exhibited in ideal MHD and hence to prevent the current-density blow-up at large times. The asymptotic result (31b) is in full quantitative agreement with the recent formulation of \textit{steady} Hall MHD near a $X$-type magnetic neutral line ([23]) which showed that (31b) indeed corresponds to a hyperbolic configuration of the magnetic field lines in the \textit{steady} case. Besides, in this range of time, the level curves of the out-of-plane magnetic field are also the streamlines of the in-plane ion flow.  

\section{Acknowledgements}

\indent I acknowledge with gratitude helpful communications and discussions with Drs. Luis Chacon, Michael Shay and Michael Johnson.

\end{document}